# WM Program manual

[1]Amani Tahat, [2]Jordi Marti, [3]Mohammad Tahat

[1,2]Department of Physics and Nuclear Engineering, Technical University of Catalonia-Barcelona Tech, Barcelona, Catalonia, Spain

[3]Department of Civil Engineering, New Mexico State University, United State of America

This user manual has been written to describe the open source code WM to be distributed associated with a research article submitted to the information technology journal **45001-ITJ-ANSI**, entitled:
"Maintenance and Reengineering of software: Creating a Visual C++ Graphical User Interface to Perform Specific Tasks Related to Soil Structure Interaction in Poroelastic Soil ".

This manual describes how to run the new produced GUI C++ program that so called 'WM' program. Section two describes the instructions of the program installation. Section three illustrates test runs description including running the program WM, sample of the input, output files, in addition to some generated graphs followed by the main form of the program created by using the Borland C++ Builder 6.

1. **Instructions OF Installation**

This program is an open source code and available on request from authors. Appendix1 presents the main form of the program created by using the Borland C++ Builder 6. At least 25 MB of free disk space is required for installing WM on user's computer (PC). The user needs to execute (SetupWM) file. Setup will create the program's shortcut in the start menu folder. The selection of different folder could be done by clicking "browse" then specifying the preferred folder. One or two additional tasks, could be selected to perform setup while installing WM, setup offers two additional tasks: (1) create a desktop icon and (2) create a quick lunch icon. After setup has been finished installing WM on user computer (PC), the application may be launched by selecting the installed icon (launch WM). But if user select the icon "create a desktop icon" then starting up the program is performed through traditional windows by clicking on an assigned icon on the desktop. The program works in a dialog mode through all steps. By choosing an item from the menu the program opens several available programs in the package. Multi windows are preserved through the execution of the program. No compilers or programming language are needed to be installed in PC.

### 3. Test Runs Description:

#### 3.1 Running the Program WM

WM is the main program in this package and it is being used for the organization of all other programs of this package. The start up of the program is performed using standard windows possibilities, for example to start using the WM program, user needs to double click on the icon "WM" after installing it, and then the WM main window will be appeared, as shown in figure 3.

Figure 3: the WM main window

The main window of the program will be opened as soon as WM program being executed. This main window contains menu items at the top of the window, the menu items include Load, CurrentVariant, and help, figure 4 presents menu structure of the WM program.

Figure 4:- menu structure of the WM package

For running the WM program, user needs to point the option " load" from the "file" menu, then a window pops up showing the default input files directory, that consists of one folder called 'samples' which includes the default input file (QQ.dat). The way of loading the input file is presented in figure 5.

Figure 5: loading the default input file from the default directory (sample).

After loading the input file, WM will produce two windows like those illustrated in Figure 6; this section provides a full description of these two windows with random screenshot to explain the way of running WM.

Figure 6: The WM main panel windows, which defines the input file and completed with default values in addition to the plotting window with its plotting menu.

### 3.1.1 WM Plotter Window

User can switch to use the WM main window, by closing the WM plotter window, as shown in figure 6. Then the massage "please, press right button mouse" will be appeared in the plotter window panel. On a mouse right click, the plotting output menu will be appeared, pointing any option will produce a specific graph based on the stored calculations that depends on the default input file (QQ.dat) which consists of two randomly variants A. Pointing the option "view tables" from the popup menu enabled viewing the output file. The options of the plotting menu are based on the mode of calculation that will be discussed in the next section .Figure 7 presents a random example of such plotting. User can continue viewing the default output by using the tabs (varianA and variantB), view log button, draw results and "recalc" in addition to draw results checkboxes.

Figure 7: Random example of using plotting window.

### 3.1.2 "WM " main window:

Figure 8 presents the main window of the program; it consists of six Input fields and three checkboxes besides the buttons (view log and variantA, variantB, variantC, "ets").

Figure 8: main window of the program WM.

The input parameters are (eta, n, kf, rhof, anus) plus "iEta" that has two values (1 and 2). Brief descriptions about the input parameters is available in table 2. Moreover, the input parameters (e.g. ieta, eta, n, rhof, anus) detailed descriptions can be found in section (2.3.2) of reference (Al Rjoub, 2007).

Table 2:- description of the input parameters.

The main window presents three checkboxes: "iSealed", "iSeepage", in addition to "Recalc & Draw results", thus, permits making multiple selections from a number of options. Normally, white space (for false) or a tick mark (for true), as illustrated in figure 8. It is clear now how the new produced GUI has linked the separated Fortran programes to be work in one station. iSeald presents the fortran code of (Todorovska and Al Rjoub, 2006a,b) and iSeepage presents the new development of this code based on (Al Rjoub, 2007).

Each checkbox has a caption help for describing the meaning of the check box. Inverting the state of a check box is done by clicking the mouse on the box. Switching to the mode of calculations can be done by using the two checkboxes ( iseapage and iseal ) easily. In the meantime, the output results can be drawn by pointing checkbox (draw result). Furthermore, user can easily view the output file by clicking the button "view log". A new checkbox will be appeared when adding a new variant, namely called ( "recalc & draw results"). One of the main purposes of the fortran code (SSI_POROUS) of (Todorovska and Al Rjoub, 2006 b) calculations was to generate data in order to draw graphs for several cases depending mainly on changing the frequency values, if the apparent building – foundation- soil system frequency changes due to water saturation the analysis begins with understanding how the wave velocities depend on frequency (eta) and on the soil permeability (for fixed value of viscosity of the fluid), which will help in, interpreting the results for the foundation stiffness and damping of (Al Rjoub, 2007). Moreover, the FORTRAN computer program was written in terms of dimensionless parameters defined using as reference: length $a$, material modulus $\mu_s$, and mass density $P_{gr}$ as explained in section 1.2 then the solution of the problem was expressed entirely in terms of dimensionless parameters, in order to control the system response (Al Rjoub, 2007).

The user of the WM program must be familiar with the theoretical background of the Fortran program (SSI_POROUS) of (Todorovska and Al Rjoub, 2006 a,b) that will simplify inserting the specified input data depending on the studying case conditions in order to generate the expected reflection coefficient (Al Rjoub, 2007).

WM program provides a very simple way for managing that. User has to start with filling his variants (A, B, C, D….etc) files and give it a comment following his studying case. WM used these variants instead of using input files in order to allow understanding the whole matter in one work station; user can save all variants in one input file, where the default input file includes two variants A and B.  Figure 9 presents the location of the (variants, comment filed) in the main window, the user comment will be appeared in the output file.

Figure 9: the fields (Variant and comment).

User has several input parameters that need to be specified. Sometimes the case study is due to fixing some parameters so that user can generate so many curves each one differs from the other by changing one parameter at each time.

For this purpose user can "delete" or "clone" more variants depending on the preconditions of the specified case study. This can be done by pointing the option "clone" or "Delete" from the "CurrentVariant " menu, as presented in figure 10. Then user can continue deleting until deletes all variants to restart the WM program by inserting the default input file QQ.dat for adding new variants.

Figure 10: the "currentVariant" menu.

**WM program provides two modes of calculation and they are:**

1. **The default mode (Al Rjoub, 2007)** : computing the coefficient of reflection for incident plane p-wave and sv-wave (P fast , P slow and S waves [coef Pf , coefPs , coefSh ] from free – half space surface when the seepage force is:

   - Included in the analysis: by clicking the mouse on the ' checkbox " 'iseapage "then to click the checkbox (recalc & draw results).
   - Not included in the analysis: user needs to repeat the previous steps but to invert the state of a check box, user must click the mouse on the checkbox '" iseapage ".

In both cases WM program will produce the WM plotter window automatically, user can close this window in order to view the output file by clicking the button "view log'.

The output window provides four services, save as, copy and paste available on the popup menu that appears on clicking the right key of the mouse. Figure 11 presents the two output files.

Figure 11: presents the two output files.

WM allows file manipulations. Thus, it provides viewing and plotting capabilities of the produced data, after inserting the variants (A, B, C, etc) and calculating the desired output file, users can generate graphic plots of data extracting from a specific output file. Figure 12 presents an example of plotting curves as those of (Al Rjoub, 2007) , to illustrate a chart of normalized wave velocities of the mixture for different values of permeability and for fixed value porosity = 0.4 and frequency (eta) = 2HZ. WM program can extract data from the stored database (output file), for plotting output data and displaying the required graphic through the WM plotter window in a matter of seconds. User can plot the two graphs one above the other at the same distance scale, with different colors to allow distinguishing them, furthermore, WM can generate several curves in the same graph not only two graphs depending on the conditions of the case study. This helps in drawing many plots with arguments that have been set the function of this program to include drawing plots and saving them as a graphic files, the user may use the right key of mouse to download the plot in clipboard, this allows pasting the plot in any data file, moreover, the program works in a dialog mode until it is being closed.

Figure 12: Example of plotting, the coefficient of reflection (Al Rjoub, 2007) for (P fast , P slow and S waves [coef Pf , coefPs , coefSh ] from free – half space surface when the seepage force is included, versus angle of incident waves of the mixture for different values of permeability and for fixed value porosity and frequency (eta),user can continue adding curves by using the plotting menu.

The status bar of the "WMplotter" window presents the values of variants (A, B, C…etc) that are located in the y axes and the values of the x axes (e.g., the angle of incident wave). The values of (X, Y) can be read by moving the curser on the specified curve, see figure 13.

Figure 13: the status bar of the window "Plotter": show current Value for nearest Graph (from cursor position).

By selecting the option "view tables" from the plotting menu, WM will generate the source file of the currently plotting data, for example if user choose to plot (coefast, coefslow and coesh) versus the angle of incident wave "I" (degree), the option view tables will generate the output file that appears in figure 14, following the specified variants, if user used 2 variants A and B then selecting the option "view tables"  will generate two different output files see figure 15.

Figure 14: source data files (output) when using variantA only.

Figure 15: the two source data files (output) when using variantA and variantB.

2. **Second mode of calculations:** involve computing the coefficient of waves reflected from a homogeneous poroelastic halfspace for incident plane p-fast wave and sealed as well as unsealed boundaries. Besides the computing of the coefficient of waves reflected from a homogeneous poroelastic halfspace for incident plane SV wave and sealed as well as unsealed boundary to be included in the same output file. That will be generated in this step all when clicking the "isealed "checkbox or not. For more clarifications, this can be done by clicking the mouse on the checkbox "isealed" for the case (sealed boundary) then to click the checkbox (recalc & draw results). User needs to repeat the previous steps to invert the state of a check box for the case (unsealed boundary) so user must click the mouse on the checkbox '" isealed'. In the two cases WM program will produce the WM plotter window automatically; user can close this window in order to view the output file by pressing the button "view log".  In this case user can continue using the "Wmplotter " window for computing:

- The reflection coefficients for incident unit displacement plane p-wave and **sealed** boundary and,

- The reflection coefficient for incident unit displacement plane SV- wave and **sealed** boundary

Herein, user has to select the option (file displace.out) from the plotting menu then the option ( cabs(uux)) to generate number of curves depending on the number of the input variants(A,B,C…etc). See figure 16.

Figure 16:- the angle of the incident wave versus (cabsuux ) for two variant (A and B).

User can view the output file by using the option "view tables" in the same way that has been discussed in the previous section. On the other hand, in order to compute the following:

- The reflection coefficients for incident unit displacement plane p-wave and **unsealed** boundary.

- The reflection coefficients for incident unit displacement plane SV-wave and **unsealed** boundary.) User has to select the option (cabs (uuz)) from the plotting window appeared in figure 16.

When checking the output file, it could be found that it consists of all expected values, that has been discussed above, the reflection coefficients appears in series way (A ,B,C, ….etc ) depending on the number of the input variants(A,B,C….etc) for example from figure 17, the C_INC_S=(1.1555887 , 0.0000000E+0) means that the reflection coefficient for incident unit displacement plane SV- wave (sealed boundary) equals to 1.1555887 and so on for all output values.

Figure17: Output file contents example.

Finally the plotting menu provides two additional options under the title ( file stresses.out ), with different submenus this part is to check the functions for ( u_r, u_th ,tau_tt.. etc ) in polar coordinates for integral of stresses along the contact surface all details are included in reference (Todorovska and Al Rjoub,2006) in addition to (Al Rjoub, 2007) study. See figure 18 and 19. User can control this from the input parameter ( iEta) that has two values (1 and 2)

- For i_eta= 1 user receives the output

  cabs (tau_zz),cabs(tau_xz),cabs(tau_xx),cabs(sigma).

- For i_eta= 2 user receives the output cabs (tau_xx), cabs(sigma) ,cabs(tau_xz).

Figure 18: How to select the option (file stress. out) menus.

Figure 19: Output, the option (file stress. out) menus.

WM generates six output files with extension "out", and they are (check, cofec1, freq_cof , freq_disp, stresses, and displace) in order to extract the plotting data following the user needs, all these files stored in the default directory (samples), in the same time.

**Figures legends**

Figure 1: The GUI of the program Adope Photoshop

Figure 2: The simple two-dimensional soil-structure interaction model

Figure 3: the WM main window

Figure 4:- menu structure of the WM package

Figure 5: loading the default input file from the default directory (sample).

Figure 6: The WM main panel windows, which defines the input file and completed with default values in addition to the plotting window with its plotting menu.

Figure 7: Random example of using plotting window.

Figure 8: main window of the program WM.

Figure 9: the fields (Variant and comment).

Figure 10: the "currentVariant" menu.

Figure11: presents the two output files.

Figure 12: Example of plotting, the coefficient of reflection for (P fast , P slow and S waves [coef Pf , coefPs , coefSh ] from free – half space surface when the seepage force is included,  versus angle of incident waves of  the mixture for different values of permeability and for fixed value porosity and  frequency (eta),user can continue adding curves by using the plotting menu

Figure 13: the status bar of the window "Plotter": show current Value for nearest Graph (from cursor position).

**Figures list**

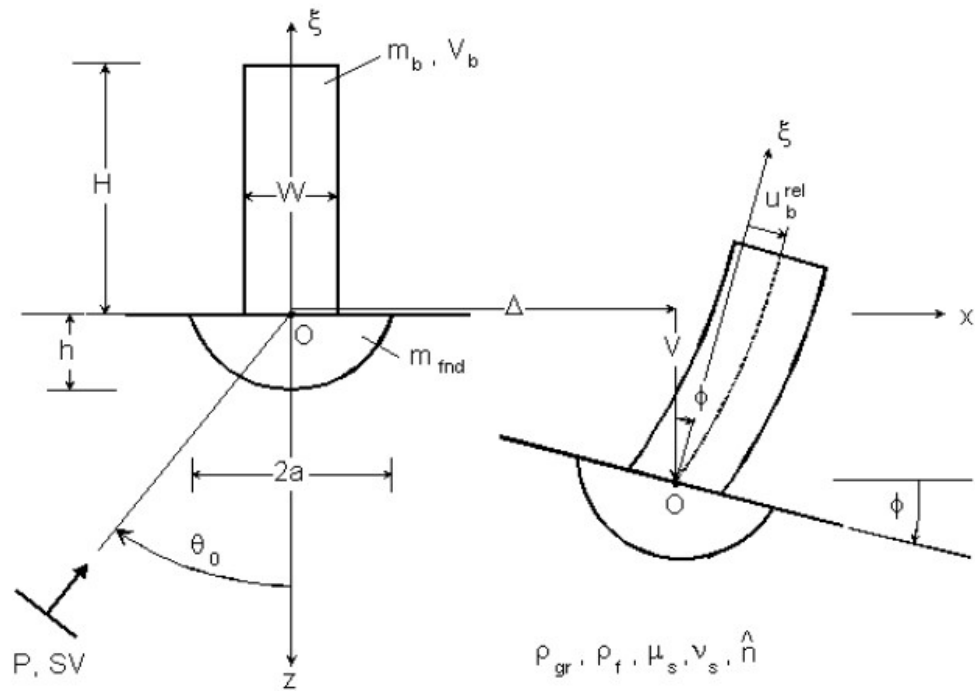

Figure 2: The simple two-dimensional soil-structure interaction model

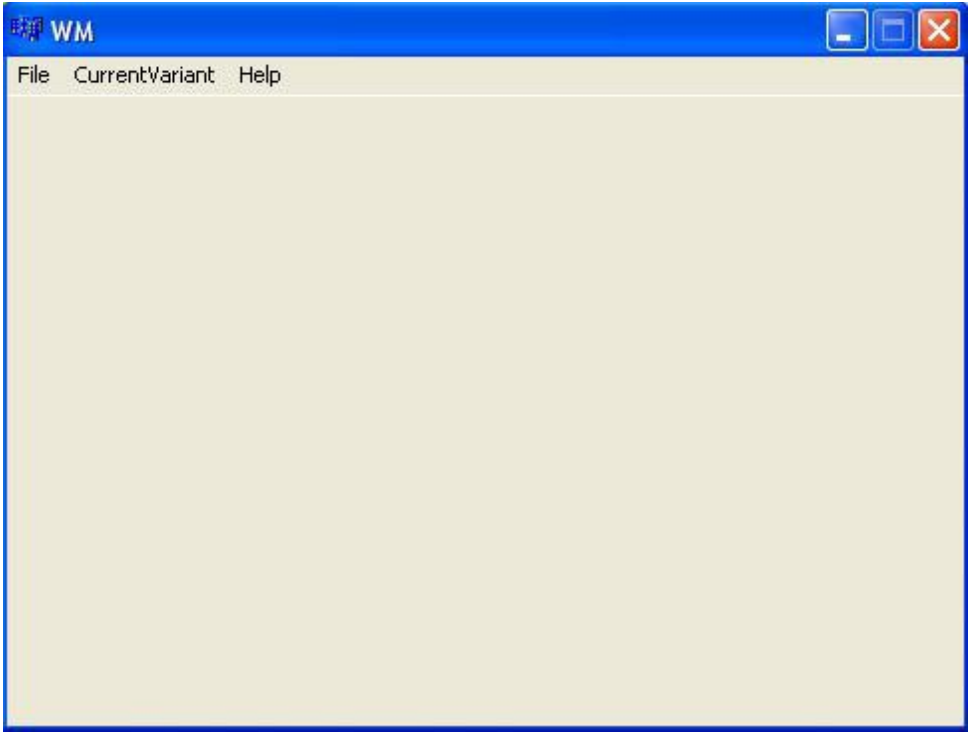

Figure 3: the WM main window

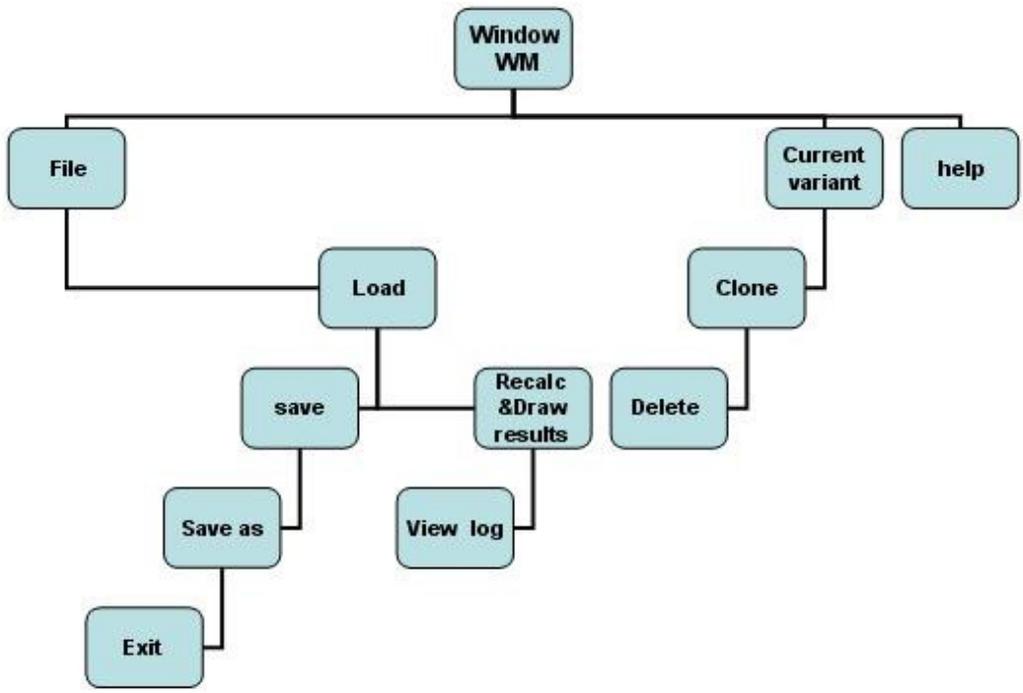

Figure 4:- menu structure of the WM package

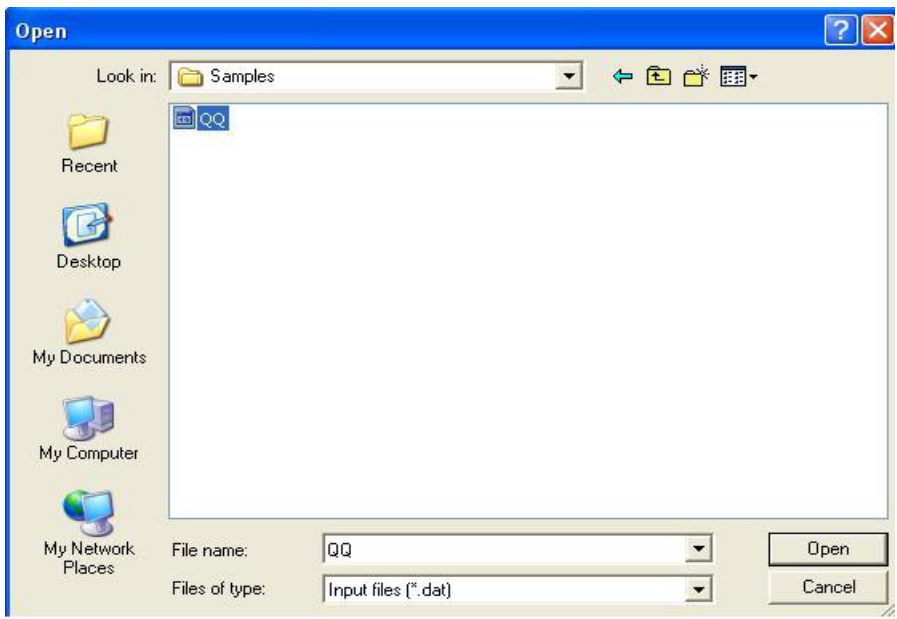

Figure 5: loading the default input file from the default directory (sample).

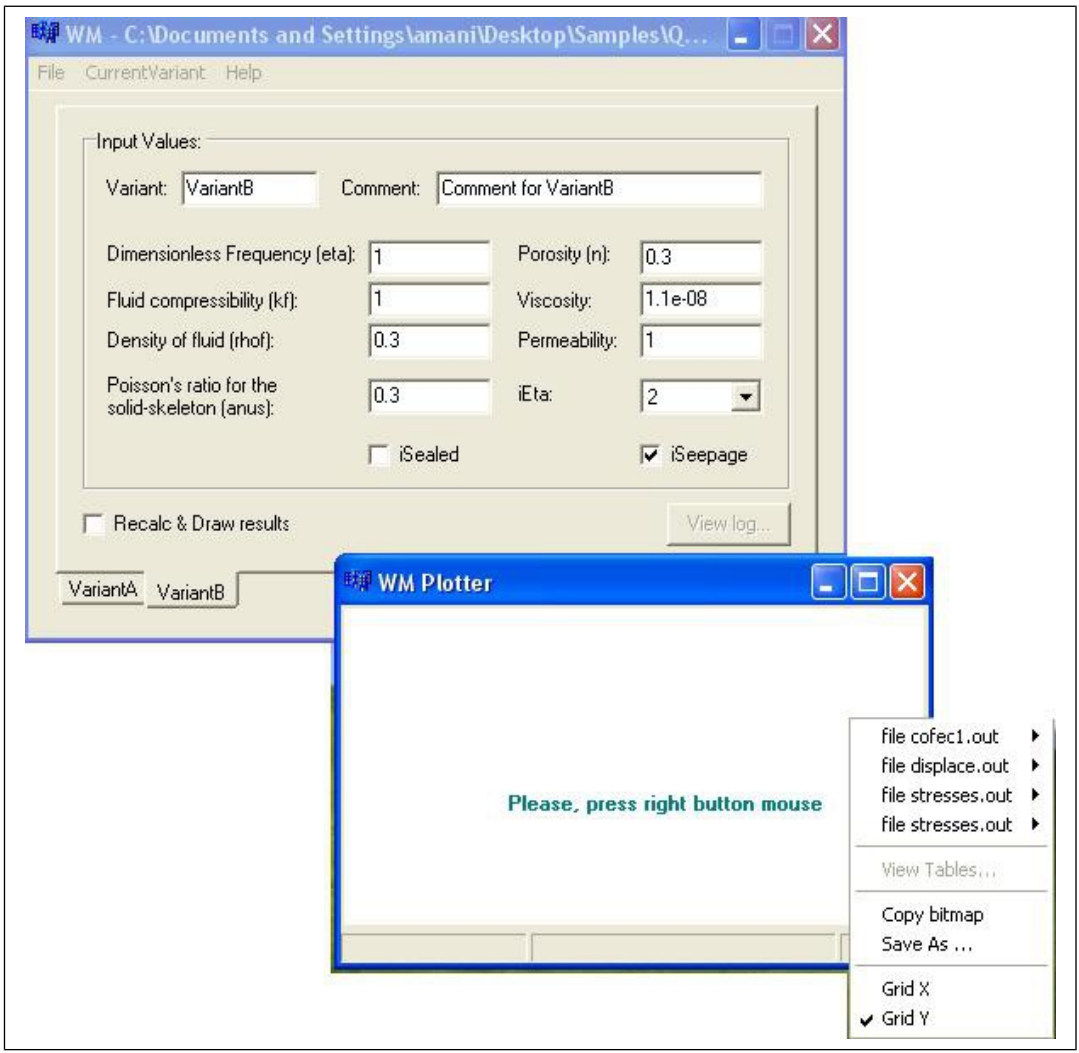

Figure 6: The WM main panel windows, which defines the input file and completed with default values in addition to the plotting window with its plotting menu.

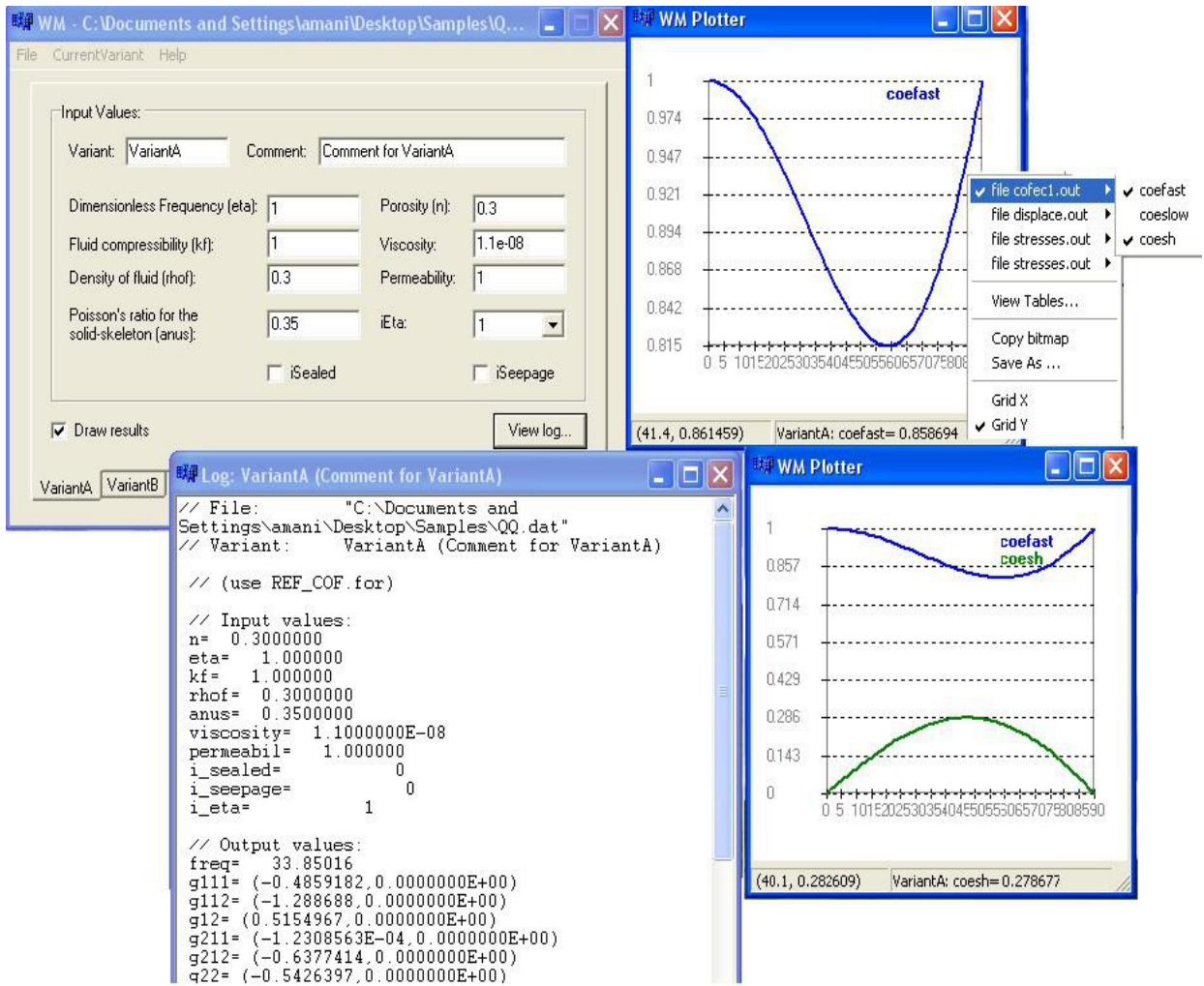

Figure 7: Random example of using plotting window.

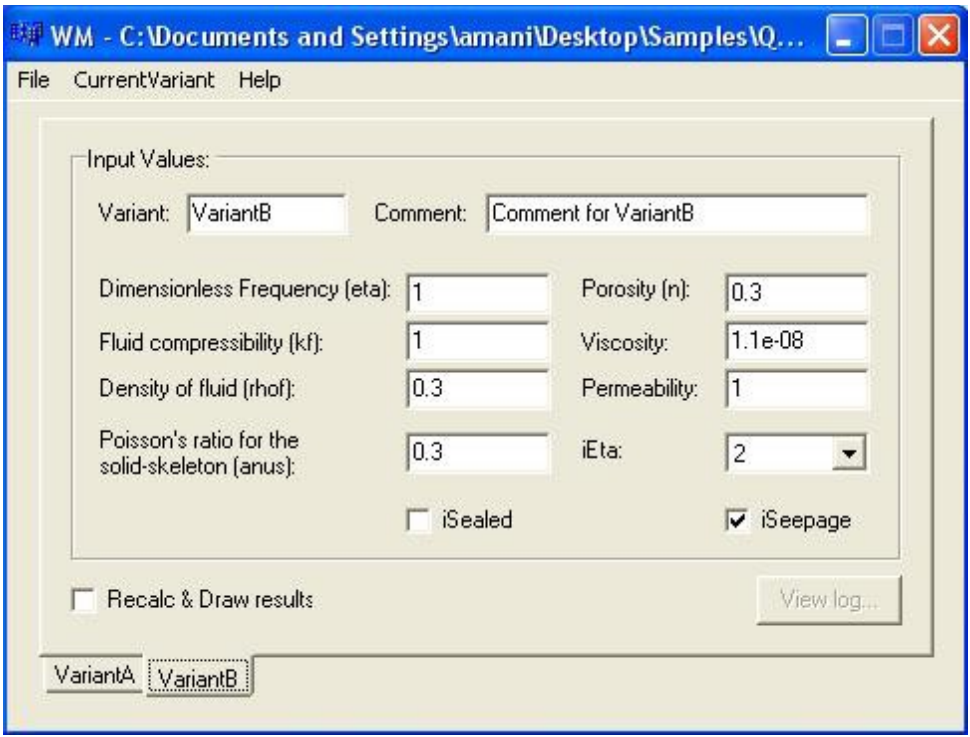

Figure 8: main window of the program WM.

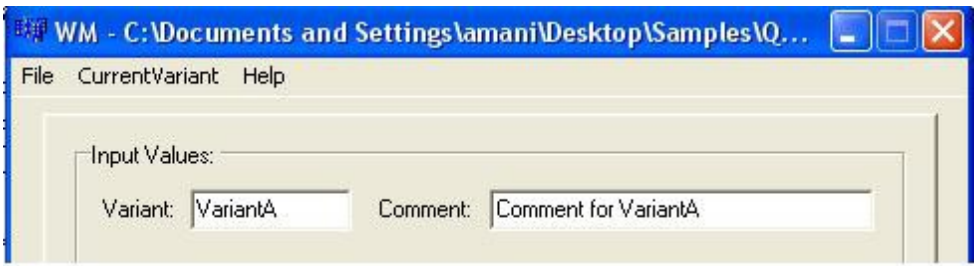

Figure 9: the fields (Variant and comment).

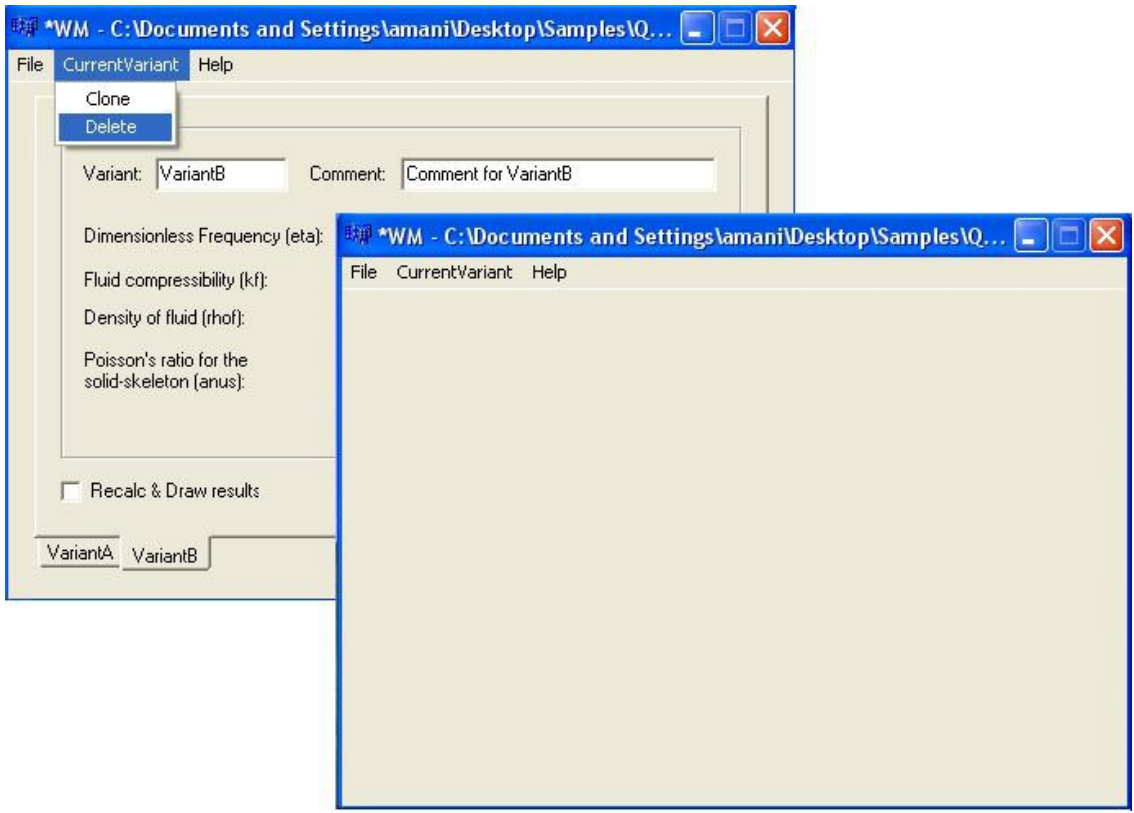

Figure 10: the "currentVariant" menu.

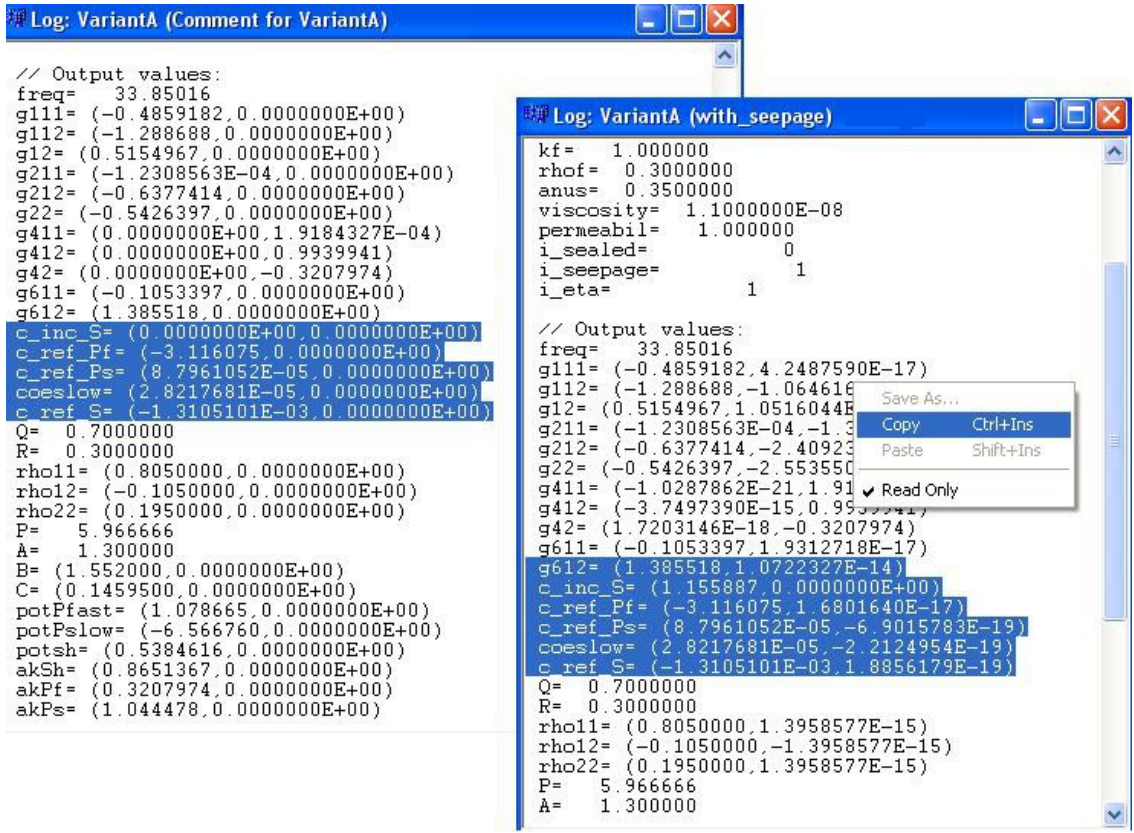

Figure11: presents the two output files.

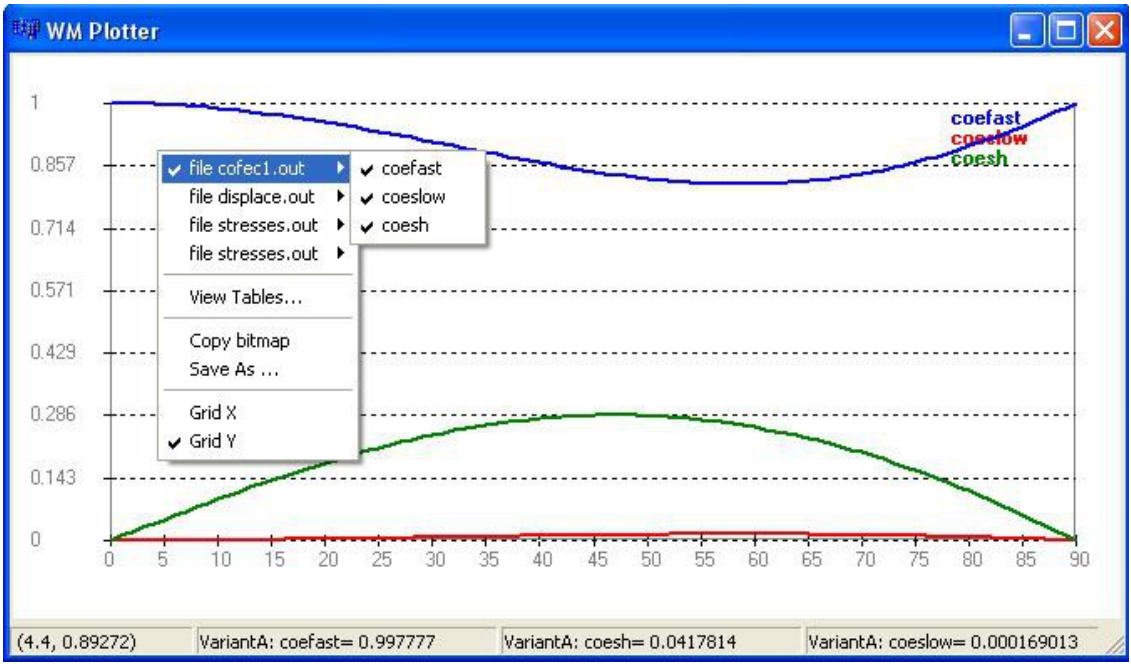

Figure 12: Example of plotting, the coefficient of reflection for (P fast , P slow and S waves [coef Pf , coefPs , coefSh ] from free – half space surface when the seepage force is included, versus angle of incident waves of the mixture for different values of permeability and for fixed value porosity and frequency (eta),user can continue adding curves by using the plotting menu.

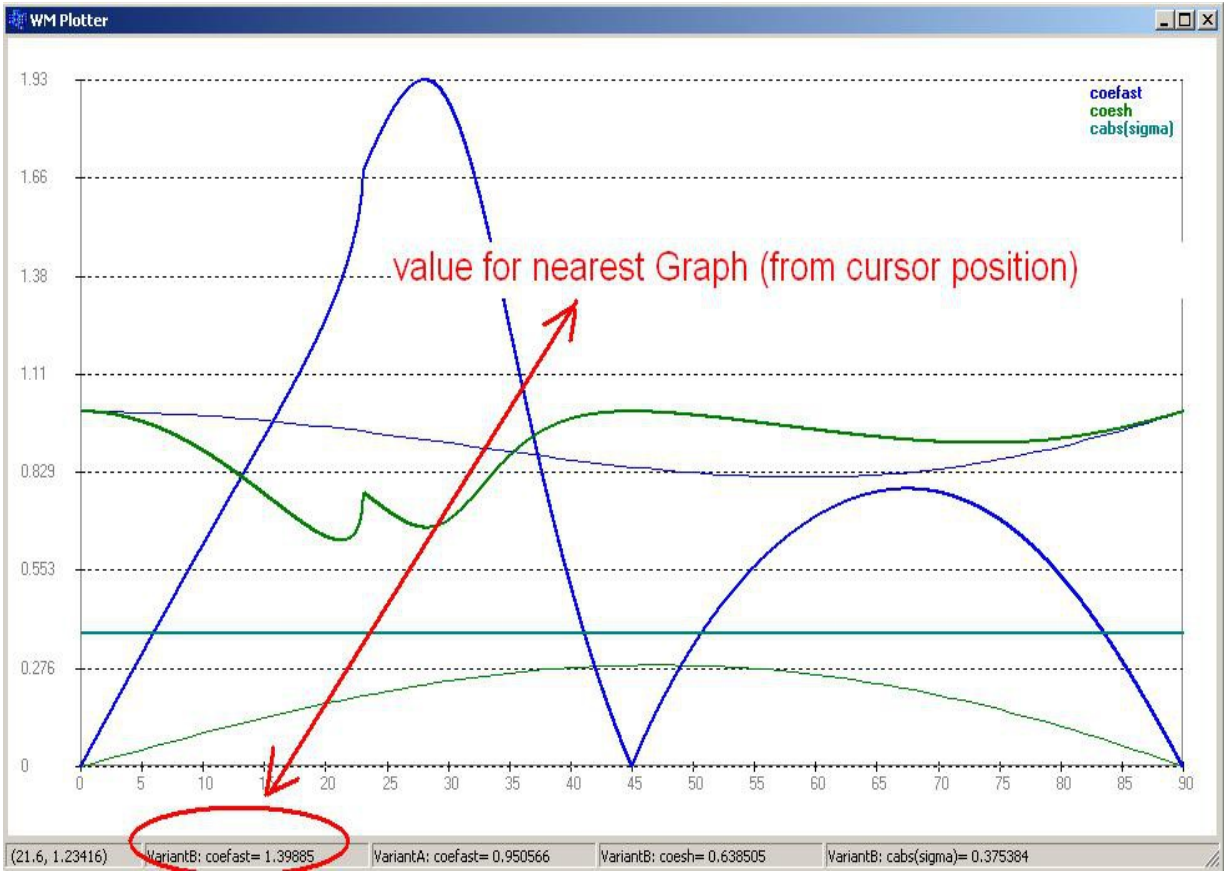

Figure 13: the status bar of the window "Plotter": show current Value for nearest Graph (from cursor position).

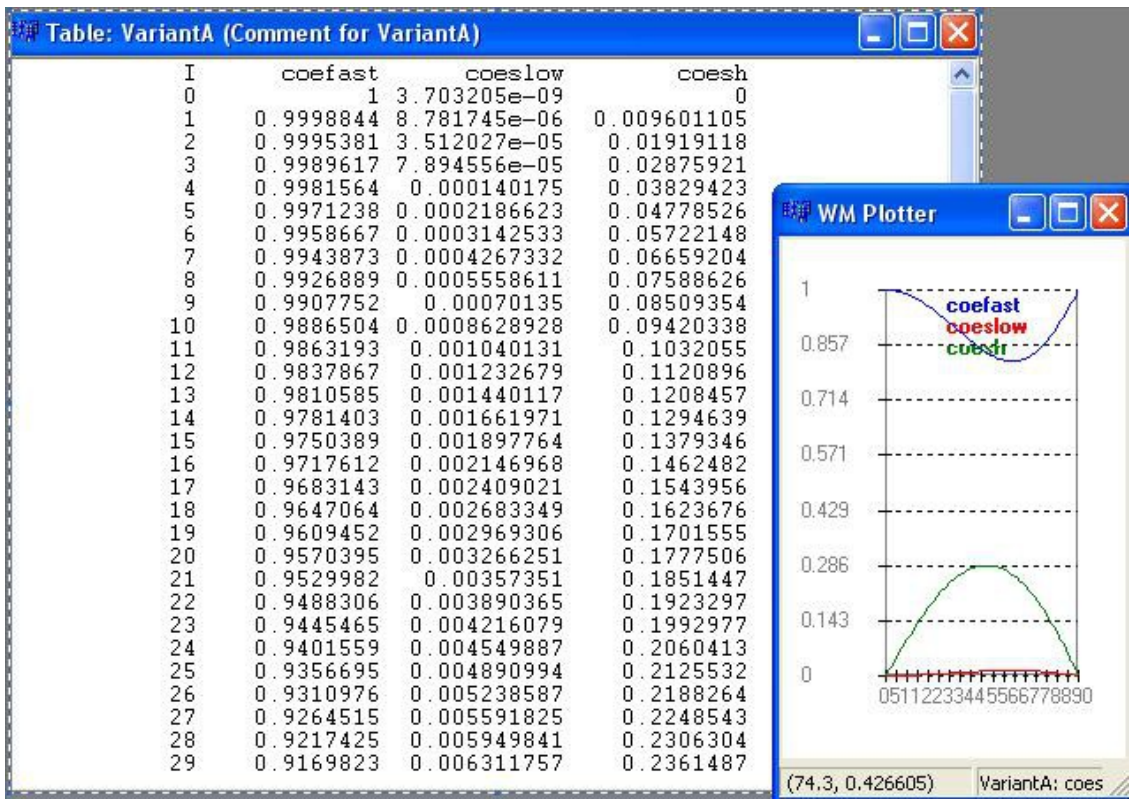

Figure 14: source data files (output) when using variantA only.

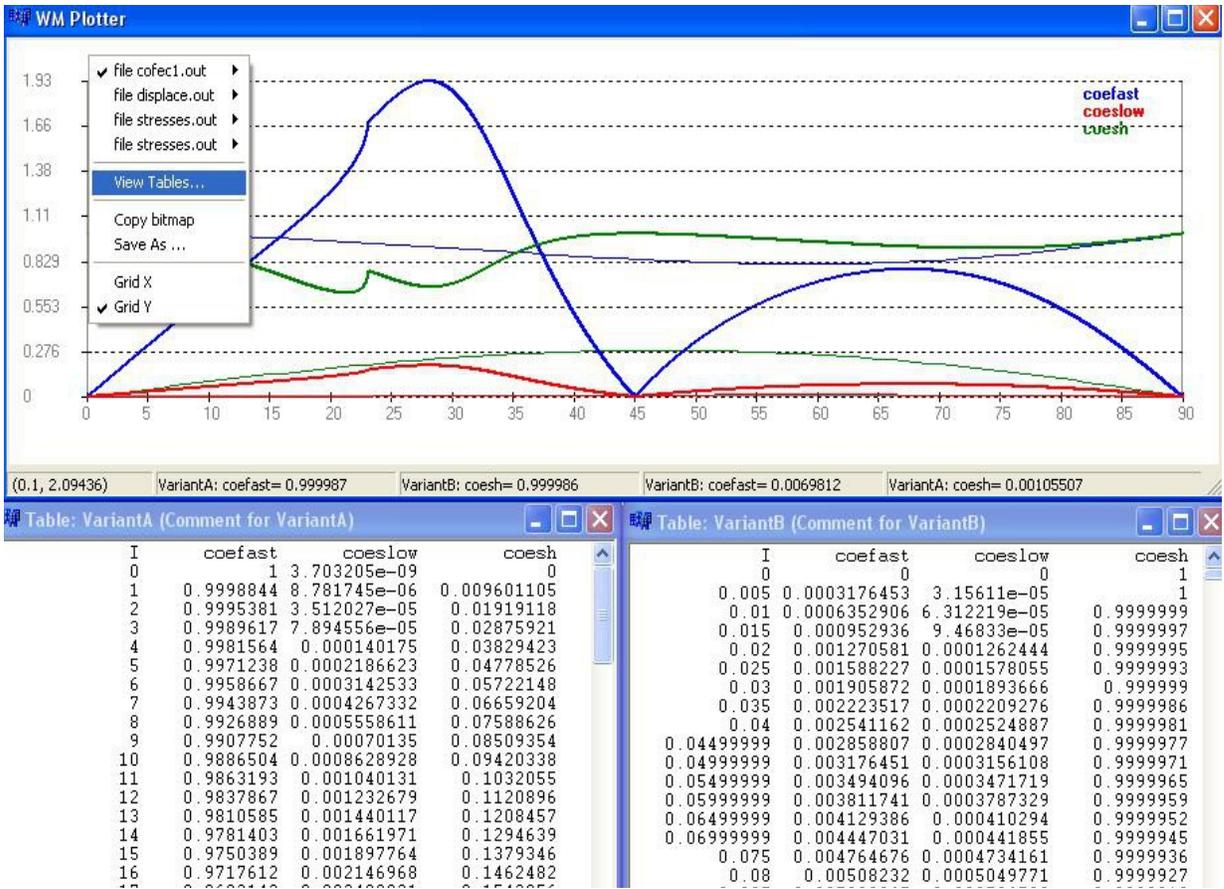

Figure 15: the two source data files (output) when using variantA and variantB.

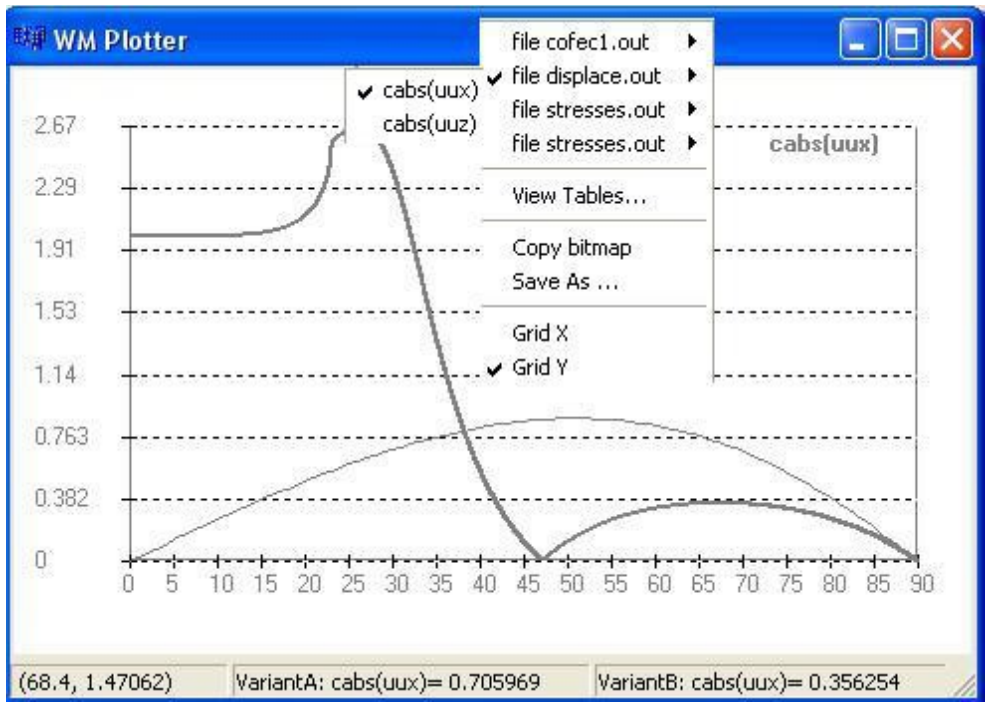

Figure 16:- the angle of the incident wave versus (cabsuux ) for two variant(A and B).

Figure17: Output file contents example.

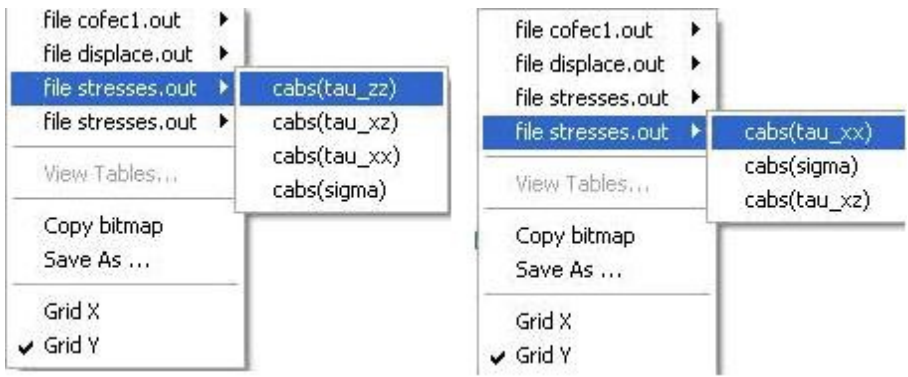

Figure 18: How to select the option (file stress. out) menus.

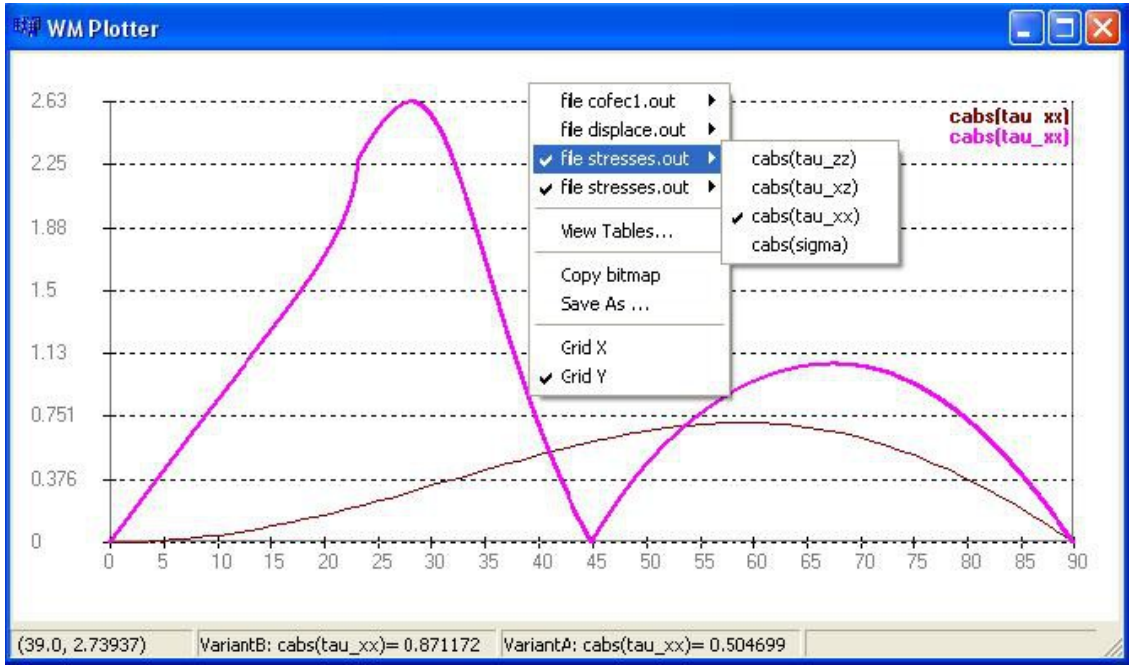

Figure 19: Output, the option (file stress. out) menus

**Appendix1: WM Main Form Created by using Borland C++ Builder 6.**

//---------------------------------------------------------------------------

#include <vcl.h>

```cpp
#include <stdio.h>

#include <process.h>

#pragma hdrstop

#include "frmMain.h"

#include "frmAbout.h"

#include "frmGraph.h"

#include "Results.h"

#include "frmViewLog.h"

//---------------------------------------------------------------------------

#pragma package(smart_init)

#pragma resource "*.dfm"

TFormMain *FormMain;

//---------------------------------------------------------------------------

TFormGraph *TFormMain::NewFormGraph() {

 if (pFormGraph) return pFormGraph;

 pFormGraph = new TFormGraph(this);

 pFormGraph->sFileVariantsFullName = ExpandFileName(sFileVariants);

return pFormGraph;

}

bool flIgnoreChangeEvent = false;
```

```cpp
//---------------------------------------------------------------------------
__fastcall TFormMain::TFormMain(TComponent* Owner) : TForm(Owner) {
 pFormGraph = NULL;
}

//---------------------------------------------------------------------------
void __fastcall TFormMain::FormCreate(TObject *Sender) {
 SetModifyVariants(false);

 OpenDialogInputVasues->InitialDir = ExtractFileDir(_argv[0]);

 InitParams();

#ifdef _DEBUG
 const char * pFileVariants = "Samples\\QQ.dat";
 VariantsLoadFromFile(pFileVariants);
#endif
}

//---------------------------------------------------------------------------
void __fastcall TFormMain::mmHelpClick(TObject *Sender) {
 TFormAbout *frm = new TFormAbout(this);
 frm->ShowModal();
 delete frm;
}
```

```cpp
//-------------------------------------------------------------------

struct ParamInfo {

 const char * pParamName;

 void * pParamControl;

 int sms;

};

//-------------------------------------------------------------------

class InputValues {

public:

 InputValues();

~InputValues() {

        delete pResults;

 }

 AnsiString sVariantIdent;

 AnsiString sVariantComment;

 double n;

 double eta;

 double kf;

 double rhof;

 double anus;
```

```cpp
    double viscosity;
//? double j;
    double permeabil;

    int i_sealed,i_seepage,i_eta;
    int iDrawGraph;

    int GetSMS(void * pMember) const { return (char*)pMember - (char*)this; }

    void SetAsDouble(int sms, const AnsiString &s) {
        double *p = (double *) (((char*)this) + sms);
        sscanf(s.c_str(),"%lg",p);
    }
    void SetAsCheckBox(int sms, const AnsiString &s) {
        int *p = (int*) (((char*)this) + sms);
        sscanf(s.c_str(),"%d",p);
    }
    void SetAsComboBox(int sms, const AnsiString &s) {
        int *p = (int*) (((char*)this) + sms);
        sscanf(s.c_str(),"%d",p);
    }

    void ToDialogParamDouble  (const ParamInfo *piEdt) const;
    void ToDialogParamDoubleAll    (const ParamInfo *piEdt) const;
    void ToDialogParamCheckBox     (const ParamInfo *piChB) const;
```

```
void ToDialogParamCheckBoxAll   (const ParamInfo *piChB) const;
void ToDialogParamComboBox      (const ParamInfo *piCoB) const;
void ToDialogParamComboBoxAll   (const ParamInfo *piCoB) const;

void FromDialogParamDouble      (const ParamInfo *piEdt);
void FromDialogParamDoubleAll   (const ParamInfo *piEdt);
void FromDialogParamCheckBox    (const ParamInfo *piChB);
void FromDialogParamCheckBoxAll(const ParamInfo *piChB);
void FromDialogParamComboBox    (const ParamInfo *piCoB);
void FromDialogParamComboBoxAll(const ParamInfo *piCoB);

void SaveVariant(FILE *out) const;
int CreateTempInputFile(const char *fnameIn, int mode, AnsiString &fnameInTmp) const;

Results *pResults;

}; //end class InputValues

//---------------------------------------------------------------------------
InputValues::InputValues() {
    n    =0.3;
    eta  =1;
    kf   =1;
    rhof =0.3;
```

```
 anus    =0.3;

 viscosity = 1e-8;
// j = 1;
 permeabil = 1;

 i_sealed      = 0;
 i_seepage     = 1;
 i_eta         = 0;

 iDrawGraph  = 0;

 pResults = NULL;
} //end InputValues::InputValues

//----------------------------------------------------------------------
void InputValues::SaveVariant(FILE *out) const {
 fprintf(out, "\n//------------------------------------------");
 fprintf(out, "\nVariantIdent=%s", sVariantIdent.c_str());
 fprintf(out, "\nVariantComment=%s", sVariantComment.c_str());

 fprintf(out, "\neta=%lg"      , eta);
 fprintf(out, "\nkf=%lg"       , kf);
 fprintf(out, "\nrhof=%lg"     , rhof);
 fprintf(out, "\nanus=%lg"     , anus);
```

```
    fprintf(out, "\nn=%lg"          , n);
    fprintf(out, "\nviscosity=%lg"  , viscosity);
    fprintf(out, "\npermeabil=%lg"  , permeabil);
//?fprintf(out, "\nj=%lg"           , j);

    fprintf(out, "\ni_sealed=%d" , i_sealed);
    fprintf(out, "\ni_seepage=%d"    , i_seepage);
    fprintf(out, "\ni_eta=%d"    , i_eta);

    fprintf(out, "\n");
    fprintf(out, "\niDrawGraph=%d"   , iDrawGraph);

    fprintf(out, "\n");
} //end InputValues::SaveVariant

//------------------------------------------------------------------------
int InputValues::CreateTempInputFile(const char * fnameIn, int mode, AnsiString &fnameInTmp) const {
    if (mode<1 || mode>2) return 1;

    const char *modeName[] = {NULL,"~SEE-REF.txt","~REF_COF.txt"};
    fnameInTmp = ChangeFileExt(fnameIn,modeName[mode]);

    FILE *out = fopen(fnameInTmp.c_str(),"wt");
```

```c
    if (!out) return 2;

    if (mode==1) {
        fprintf(out, "%lg  %lg  %lg  %lg\n", n, kf, rhof, anus);
        fprintf(out, "%lg ", viscosity);
        //fprintf(out, " %lg\n", j);
        fprintf(out, "%d %d %d\n", i_sealed, i_seepage, i_eta);
        fprintf(out, "\n");
        fprintf(out, "\n// n, kf, rhof, anus");
        fprintf(out, "\n// viscosity, j");
        fprintf(out, "\n// i_sealed, i_seepage, i_eta");
    }
    else
    if (mode==2) {
        fprintf(out, "%lg  %lg  %lg  %lg  %lg\n", n, eta, kf, rhof, anus);
        fprintf(out, "%lg  %lg\n", viscosity, permeabil);
        fprintf(out, "%d %d %d\n", i_sealed, i_seepage, i_eta);
        fprintf(out, "\n");
        fprintf(out, "\n// n, eta, kf, rhof, anus");
        fprintf(out, "\n// viscosity, permeabil");
        fprintf(out, "\n// i_sealed, i_seepage, i_eta");
    }

    fclose(out);
    return 0;
```

} //end InputValues::CreateTempInputFile

//-----------------------------------------------------------------------
```cpp
void InputValues::ToDialogParamDouble(const ParamInfo *piEdt) const {
 double *p = (double *) (((char*)this) + piEdt->sms);
 AnsiString s;
 s.printf("%lg",*p);
 TEdit *pEdt = (TEdit *) piEdt->pParamControl;
 pEdt->Text = s;
}
void InputValues::ToDialogParamDoubleAll(const ParamInfo *piEdt) const {
 for(; piEdt->pParamName; piEdt++) {
        ToDialogParamDouble(piEdt);
 }
}
```
//-----------------------------------------------------------------------
```cpp
void InputValues::ToDialogParamCheckBox(const ParamInfo *piChB) const {
 int *p = (int *) (((char*)this) + piChB->sms);
 TCheckBox *pCheckBox = (TCheckBox *) piChB->pParamControl;
 pCheckBox->Checked = *p;
}
void InputValues::ToDialogParamCheckBoxAll(const ParamInfo *piChB) const {
 for(; piChB->pParamName; piChB++) {
        ToDialogParamCheckBox(piChB);
 }
```

```cpp
}
//---------------------------------------------------------------------------
void InputValues::ToDialogParamComboBox(const ParamInfo *piCoB) const {
    int *p = (int *) (((char*)this) + piCoB->sms);
    AnsiString sp = IntToStr(*p);
    TComboBox *pComboBox = (TComboBox*) piCoB->pParamControl;
    for(int i=0; i<pComboBox->Items->Count; i++) {
        String ss = pComboBox->Items->Strings[i];
        if (ss==sp) {
            pComboBox->ItemIndex = i;
            return;
        }
    }
    pComboBox->ItemIndex = -1;
}
void InputValues::ToDialogParamComboBoxAll(const ParamInfo *piCoB) const {
    for(; piCoB->pParamName; piCoB++) {
        ToDialogParamComboBox(piCoB);
    }
}

//---------------------------------------------------------------------------
void InputValues::FromDialogParamDouble(const ParamInfo *piEdt) {
    TEdit *pEdt = (TEdit *) piEdt->pParamControl;
```

```cpp
 AnsiString s = pEdt->Text;

 double *p = (double *) (((char*)this) + piEdt->sms);

 sscanf(s.c_str(),"%lg",p);

}
//---------------------------------------------------------------------------
void InputValues::FromDialogParamDoubleAll(const ParamInfo *piEdt) {

 for(; piEdt->pParamName; piEdt++)

        FromDialogParamDouble(piEdt);

}

//---------------------------------------------------------------------------
void InputValues::FromDialogParamCheckBox(const ParamInfo *piChB) {

 TCheckBox *pCheckBox = (TCheckBox *) piChB->pParamControl;

 int *p = (int *) (((char*)this) + piChB->sms);

 *p = pCheckBox->Checked;

}
//---------------------------------------------------------------------------
void InputValues::FromDialogParamCheckBoxAll(const ParamInfo *piChB) {

 for(; piChB->pParamName; piChB++)

        FromDialogParamCheckBox(piChB);

}

//---------------------------------------------------------------------------
void InputValues::FromDialogParamComboBox(const ParamInfo *piCoB) {

 TComboBox *pComboBox = (TComboBox*) piCoB->pParamControl;
```

```
    int *p = (int *) (((char*)this) + piCoB->sms);
    if (pComboBox->ItemIndex>=0)
            *p = StrToInt(pComboBox->Text);
}
//---------------------------------------------------------------------------
void InputValues::FromDialogParamComboBoxAll(const ParamInfo *piCoB) {
    for(; piCoB->pParamName; piCoB++)
            FromDialogParamComboBox(piCoB);
}

//---------------------------------------------------------------------------
ParamInfo *FindParam(ParamInfo *PI, const AnsiString &sParamName) {
    for(; PI->pParamName; PI++) {
            if (sParamName==PI->pParamName)
                    return PI;
    }
    return NULL;
}

//---------------------------------------------------------------------------
void TFormMain::ClearFields() {
    EdtPorosity              ->Text = "";
    EdtFrequencyEta          ->Text = "";
    EdtFluidCompressibilityKf->Text = "";
    EdtDensityFluidRhof      ->Text = "";
```

```
  EdtPoissonRatioAnus      ->Text = "";
  EdtViscosity          ->Text = "";
  EdtPermeability       ->Text = "";
}

//---------------------------------------------------------------------------
//---------------------------------------------------------------------------
typedef DynamicArray <ParamInfo> DynamicArrayParamInfo;
ParamInfo *ParamInfoEdt = NULL;
ParamInfo *ParamInfoChB = NULL;
ParamInfo *ParamInfoCoB = NULL;

//---------------------------------------------------------------------------
void AddParam(DynamicArrayParamInfo &daParamInfo, const char * pParamName, void * pParamControl, int ParamSms) {
 ParamInfo &pi = daParamInfo[daParamInfo.Length++];
 pi.pParamName     = strdup(pParamName);
 pi.pParamControl= pParamControl;
 pi.sms        = ParamSms;
}
//---------------------------------------------------------------------------
ParamInfo *MoveParams(DynamicArrayParamInfo &daParamInfoTmp) {
 const int nP = daParamInfoTmp.Length;
 ParamInfo *pPI = new ParamInfo[nP+1];
 ParamInfo *p = pPI;
```

```
    for(int i=0; i<nP; i++) {

        const ParamInfo &pi = daParamInfoTmp[i];

        *p++ = pi;

    }

    p->pParamName = NULL;

    p->pParamControl = NULL;

    p->sms = 0;

    daParamInfoTmp.Length = 0;

    return pPI;

} //end MoveParams

//---------------------------------------------------------------------------

void TFormMain::InitParams() {

    DynamicArrayParamInfo daParamInfoTmp;

    InputValues d;

    AddParam(daParamInfoTmp,"n"         ,EdtPorosity
        ,d.GetSMS(&d.n));

    AddParam(daParamInfoTmp,"eta"       ,EdtFrequencyEta
        ,d.GetSMS(&d.eta));

    AddParam(daParamInfoTmp,"kf"        ,EdtFluidCompressibilityKf
        ,d.GetSMS(&d.kf));

    AddParam(daParamInfoTmp,"rhof"      ,EdtDensityFluidRhof
        ,d.GetSMS(&d.rhof));

    AddParam(daParamInfoTmp,"anus"      ,EdtPoissonRatioAnus
        ,d.GetSMS(&d.anus));
```

```
    AddParam(daParamInfoTmp,"viscosity"   ,EdtViscosity
        ,d.GetSMS(&d.viscosity));

    AddParam(daParamInfoTmp,"permeabil"   ,EdtPermeability
        ,d.GetSMS(&d.permeabil));

 ParamInfoEdt = MoveParams(daParamInfoTmp);

    AddParam(daParamInfoTmp,"i_sealed"    ,CheckBoxSealed
        ,d.GetSMS(&d.i_sealed));

    AddParam(daParamInfoTmp,"i_seepage"   ,CheckBoxSeepage
        ,d.GetSMS(&d.i_seepage));

    AddParam(daParamInfoTmp,"iDrawGraph",CheckBoxDrawGraph
        ,d.GetSMS(&d.iDrawGraph));

 ParamInfoChB = MoveParams(daParamInfoTmp);

    AddParam(daParamInfoTmp,"i_eta",ComboBoxIEta
        ,d.GetSMS(&d.i_eta));

 ParamInfoCoB = MoveParams(daParamInfoTmp);

} //end TFormMain::InitParams

//---------------------------------------------------------------------
void TFormMain::VariantsLoadFromFile(const char * pFileVariants) {

 if (TabControllVariants->Tabs->Count<=0) SetModifyVariants(false);

 FILE *in = fopen(pFileVariants,"rt");
 if (!in) {
```

```c
        sFileVariants = "";
        return;
}

InputValues *pIV = NULL;

char buf[1000];
for(;;) {
        char * p = fgets(buf, sizeof(buf)-1, in);
        if (!p) break;

        while(*p==' ' || *p=='\t') p++;
        if (*p=='/') continue;   // - comment

        char *e = p;
        for(; *e; e++) {
                if (strchr(" \t=",*e)) break;
        }
        if (*e=='\0') continue;

        char *v;
        if (*e=='=') v = e+1;
        else {
                v = strchr(e,'=');
                if (!v) continue;
```

```
                v++;
        }
        *e = '\0';

        AnsiString sParamName     = AnsiString(p).Trim();
        AnsiString sParamValue    = AnsiString(v).Trim();

        if (sParamName=="VariantIdent") {
                pIV = new InputValues();
                pIV->sVariantIdent = sParamValue;
                AddTab(pIV);
                continue;
        }

        if (sParamName=="VariantComment") {
                pIV->sVariantComment = sParamValue;
                continue;
        }

//if (sParamName=="n") {
// String as="";
//}
        if (!pIV) continue;
        ParamInfo *pPI = FindParam(ParamInfoEdt, sParamName);
        if (pPI) {
```

```
                pIV->SetAsDouble(pPI->sms, sParamValue);
                //pIV->ToDialogParamDouble1(pPI);
                continue;
        }
        pPI = FindParam(ParamInfoChB, sParamName);
        if (pPI) {
                pIV->SetAsCheckBox(pPI->sms, sParamValue);
                continue;
        }
        pPI = FindParam(ParamInfoCoB, sParamName);
        if (pPI) {
                pIV->SetAsComboBox(pPI->sms, sParamValue);
                continue;
        }

        //? Unknown Param

} //end for
fclose(in);

sFileVariants = pFileVariants;
Caption = AnsiString("WM - ") + sFileVariants;

// Draw On LoadFile variants
int isDraw = 0;
```

```cpp
    const int nV = TabControllVariants->Tabs->Count;
    for(int i=0; i<nV; i++) {
        InputValues *pIV = (InputValues *) TabControllVariants->Tabs->Objects[i];
        if (!pIV->iDrawGraph) continue;
        if (!pIV->pResults) Recalc(pIV);
        isDraw |= (int) pIV->pResults;
    } //end for
    if (isDraw) {
        if (!pFormGraph) pFormGraph = NewFormGraph(); // new TFormGraph(this);
        pFormGraph->UpdateGraphData();
        pFormGraph->BringToFront();
    }

    InputValuesSetToForm(pIV);

} //end TFormMain::VariantsLoadFromFile

//---------------------------------------------------------------------------
void TFormMain::VariantSaveToFile(const char * pFileVariants) {
 int nV = TabControllVariants->Tabs->Count;
 if (nV<=0) return;

 AnsiString sFileBAK = ChangeFileExt(AnsiString(pFileVariants),".bak");
 unlink(sFileBAK.c_str());
 rename(pFileVariants,sFileBAK.c_str());
```

```cpp
    FILE *out = fopen(pFileVariants,"wt");
    if (!out) return;
    for(int i=0; i<nV; i++) {
        const InputValues *pIV = (const InputValues *) TabControllVariants->Tabs->Objects[i];
        pIV->SaveVariant(out);
    } //end for
    fclose(out);

    SetModifyVariants(false);
    sFileVariants = pFileVariants;
} //end TFormMain::VariantSaveToFile

//---------------------------------------------------------------------------
void TFormMain::InputValuesSetToForm(const InputValues *pIV) {
    if (!pIV) {
        // ???
        ClearFields();
        return;
    }
    flIgnoreChangeEvent = true;

    EdtVariantIdent     ->Text = pIV->sVariantIdent;
    EdtVariantComment  ->Text = pIV->sVariantComment;
```

```
 pIV->ToDialogParamDoubleAll    (ParamInfoEdt);

 pIV->ToDialogParamCheckBoxAll (ParamInfoChB);

 pIV->ToDialogParamComboBox    (ParamInfoCoB);

 SetCaptions(pIV->pResults);

 flIgnoreChangeEvent = false;
} //end TFormMain::InputValuesSetToForm
//---------------------------------------------------------------
void TFormMain::InputValuesGetFromForm(InputValues *pIV) {
 if (!pIV) return;
 pIV->sVariantIdent   = EdtVariantIdent    ->Text;
 pIV->sVariantComment = EdtVariantComment  ->Text;
 pIV->FromDialogParamDoubleAll  (ParamInfoEdt);
 pIV->FromDialogParamCheckBoxAll(ParamInfoChB);
 pIV->FromDialogParamComboBoxAll(ParamInfoCoB);
} //end TFormMain::InputValuesGetFromForm

//---------------------------------------------------------------
void TFormMain::AddTab(InputValues *pIV) {
 TabControllVariants->Tabs->Add(pIV->sVariantIdent);
 int TabIndex = TabControllVariants->Tabs->Count-1;
 TabControllVariants->Tabs->Objects[TabIndex] = (TObject*) pIV;
 SetTabIndex(TabIndex);
```

```
}

//---------------------------------------------------------------------------
void TFormMain::SetTabIndex(int TabIndex) {
  TabControllVariants->Visible = TabControllVariants->Tabs->Count > 0;
  if (TabIndex>=0) {
       TabControllVariants->TabIndex = TabIndex;
       const InputValues *pIV = (const InputValues *) TabControllVariants->Tabs->Objects[TabIndex];
       InputValuesSetToForm(pIV);
  } else {
       TabControllVariants->TabIndex = -1;
       ClearFields();
  }
}

//---------------------------------------------------------------------------
void __fastcall TFormMain::TabControllVariantsChanging(TObject *Sender, bool &AllowChange) {
  if (TabControllVariants->TabIndex<0) return;
  InputValues *pIV = (InputValues *) TabControllVariants->Tabs->Objects[TabControllVariants->TabIndex];
  InputValuesGetFromForm(pIV);
// flIgnoreChangeEvent = true;
}
//---------------------------------------------------------------------------
```

```cpp
void __fastcall TFormMain::TabControllVariantsChange(TObject *Sender) {
 SetTabIndex(TabControllVariants->TabIndex);
// flIgnoreChangeEvent = false;
 if (pFormGraph) {
        pFormGraph->Invalidate();
        if (CheckBoxDrawGraph->Checked) pFormGraph->Show();
        this->BringToFront();
 }
}

//---------------------------------------------------------------------------
void __fastcall TFormMain::mmCurrentVariantClick(TObject *Sender) {
 bool en = TabControllVariants->TabIndex >= 0;
 mmCurrentVariantDelete    ->Enabled = en;
 mmCurrentVariantClone     ->Enabled = en;
}
//---------------------------------------------------------------------------
void __fastcall TFormMain::mmCurrentVariantCloneClick(TObject *Sender) {
 if (TabControllVariants->TabIndex<0) return;
 const InputValues *pIVSrc = (const InputValues *) TabControllVariants->Tabs->Objects[TabControllVariants->TabIndex];
 InputValues *pIVDest = new InputValues();
 *pIVDest = *pIVSrc;
 pIVDest->pResults = NULL;
 pIVDest->sVariantIdent += "~Clone";
```

```cpp
  AddTab(pIVDest);

  SetModifyVariants(true);

}
//---------------------------------------------------------------------------
void __fastcall TFormMain::mmCurrentVariantDeleteClick(TObject *Sender) {

 // TabControllVariants->Tabs->Clear();

 int TabIndex = TabControllVariants->TabIndex;

 if (TabIndex<0) return;

 InputValues *pIV = (InputValues *) TabControllVariants->Tabs->Objects[TabIndex];

 TabControllVariants->Tabs->Delete(TabIndex);

 delete pIV;

 if (TabIndex>=TabControllVariants->Tabs->Count) TabIndex = TabControllVariants->Tabs->Count-1;

 SetTabIndex(TabIndex);

 SetModifyVariants(true);

}

//---------------------------------------------------------------------------
void __fastcall TFormMain::mmLoadClick(TObject *Sender) {

 OpenDialogInputVasues->Filter = "Input files (*.dat)|*.dat|All files (*.*)|*.*|";

 if (!OpenDialogInputVasues->Execute()) return;

 VariantsLoadFromFile(OpenDialogInputVasues->FileName.c_str());

}
```

```cpp
//---------------------------------------------------------------------------
void __fastcall TFormMain::mmExitClick(TObject *Sender) {
 Close();
}

//---------------------------------------------------------------------------
void __fastcall TFormMain::File1Click(TObject *Sender) {
 bool enSave = isModifyVariants && !sFileVariants.IsEmpty() && TabControllVariants->Tabs->Count>0;
 mmSave->Enabled = enSave;

 bool enSaveAs = TabControllVariants->Tabs->Count>0;
 mmSaveAs->Enabled = enSaveAs;
}
//---------------------------------------------------------------------------
void __fastcall TFormMain::mmSaveClick(TObject *Sender) {
 VariantSaveToFile(sFileVariants.c_str());
}
//---------------------------------------------------------------------------
void __fastcall TFormMain::mmSaveAsClick(TObject *Sender) {
 SaveDialogInputVasues->Filter = "Input files (*.dat)|*.dat|All files (*.*)|*.*|";
 if (!SaveDialogInputVasues->Execute()) return;
 VariantSaveToFile(SaveDialogInputVasues->FileName.c_str());
}
```

```cpp
//---------------------------------------------------------------------------
void __fastcall TFormMain::EdtVariantIdentChange(TObject *Sender) {
 if (TabControllVariants->TabIndex<0) return;
 TabControllVariants->Tabs->Strings[TabControllVariants->TabIndex] = EdtVariantIdent->Text;
 if (!flIgnoreChangeEvent) SetModifyVariants(true);
}
//---------------------------------------------------------------------------
void __fastcall TFormMain::EdtVariantCommentChange(TObject *Sender) {
 if (!flIgnoreChangeEvent) SetModifyVariants(true);
}

//---------------------------------------------------------------------------
void __fastcall TFormMain::OnChangeInputValues(TObject *Sender) {
 if (flIgnoreChangeEvent) return;
 if (TabControllVariants->TabIndex<0) return;
 SetModifyVariants(true);

 InputValues *pIV = (InputValues *) TabControllVariants->Tabs->Objects[TabControllVariants->TabIndex];
 if (pIV->pResults) {
        delete pIV->pResults;
        pIV->pResults = NULL;
        //? if (pIV->iDrawGraph) pIV->iDrawGraph = 0;
        //? if (pFormGraph) pFormGraph->Invalidate();
 }
```

```
  SetCaptions(pIV->pResults);

}

//---------------------------------------------------------------

typedef int (*tdCalculation)(int mode, char * fnameInTmp, char * fnameOut, char * folderOut);

tdCalculation pFunCalculation = NULL;

//---------------------------------------------------------------

void GetProcAddresses() {

 static HMODULE hModuleDll = NULL;

 if (hModuleDll) return;

 AnsiString sDllFullName = ExtractFileDir(_argv[0]) + "\\WM.dll";

 hModuleDll = ::LoadLibrary(sDllFullName.c_str());

 pFunCalculation = (tdCalculation) GetProcAddress(hModuleDll, "Calculation");

}

//---------------------------------------------------------------

AnsiString LoadTextFile(const char * pFileName) {

 FILE *in = fopen(pFileName,"rt");

 if (!in) return "Error !";

 AnsiString s;

 char buf[1000];

 for(;;) {

        char * p = fgets(buf, sizeof(buf)-1, in);

        if (!p) break;
```

```
              s += AnsiString(p).TrimRight() + "\r\n";
       }
       fclose(in);
       return s;
}

//---------------------------------------------------------------------------
void TFormMain::Recalc(InputValues *pIV) {
       static int nv=1;
       if (nv) {
              nv=0;
              GetProcAddresses();
              if (!pFunCalculation) {
                     Application->MessageBoxA("Couldn't load DLL", NULL, MB_OK);
                     return;
              }
       }
       if (!pFunCalculation) return;

       AnsiString sInputFileName = sFileVariants;
       if (sInputFileName.IsEmpty()) sInputFileName = _argv[0];
       sInputFileName = ExpandFileName(sInputFileName);

       AnsiString fnameOut = ChangeFileExt(sInputFileName,"~Log.txt");
       AnsiString folderOut = ExtractFileDir(fnameOut) + "\\";
```

```
AnsiString fnameInTmp;
int mode = 2;
int ret = pIV->CreateTempInputFile(sInputFileName.c_str(), mode, fnameInTmp);
if (ret) {
    AnsiString msg;
    msg.printf("Error N %d on Create Temp Input File\r\n\"%s\"", ret, fnameInTmp.c_str());
    MessageBox(this, msg.c_str(), NULL, MB_OK);
    return;
}

Screen->Cursor = crHourGlass;
Application->ProcessMessages();

try {
    pFunCalculation(mode, fnameInTmp.c_str(), fnameOut.c_str(), folderOut.c_str());
} catch(...) {
}

// VerifyButtonsEnabled();
// OnBtnViewOutputLogFile();

delete pIV->pResults;
Results *pResults = pIV->pResults = new Results();
```
56

```
    pResults->nPoints =
 ReadGraphData(folderOut+"cofec1.out"    , &pResults->pX , &pResults->pC1, &pResults->pC2, &pResults->pC3, NULL);
 ReadGraphData(folderOut+"displace.out"    , NULL, &pResults->pD1, &pResults->pD2, NULL, NULL);

 if (pIV->i_eta==1) {
        ReadGraphData(folderOut+"stresses.out"    , NULL, &pResults->pS1, &pResults->pS2, &pResults->pS3, &pResults->pS4);
 } else {
        ReadGraphData(folderOut+"stresses.out"    , NULL, &pResults->pS31, &pResults->pS32, &pResults->pS33, NULL);
 }

 pResults->sLogInfo = LoadTextFile(fnameOut.c_str());

#ifdef _DEBUG
 bool flDeleteFiles = false;
// flDeleteFiles = true;
#else
 bool flDeleteFiles = true;
#endif
 if (flDeleteFiles) {
        unlink(fnameInTmp.c_str());
        unlink(fnameOut.c_str());
//      DeleteFile(folderOut+"cofec1.out");
```

```
//        DeleteFile(folderOut+"displace.out");
//        DeleteFile(folderOut+"stresses.out");
//        DeleteFile(folderOut+"freq_cof.out");      // - ç
//        DeleteFile(folderOut+"freq_disp.out");
//        DeleteFile(folderOut+"check.out");
 }

 Screen->Cursor = crDefault;
 Application->ProcessMessages();

} //end TFormMain::Recalc

//---------------------------------------------------------------------------
void __fastcall TFormMain::CheckBoxDrawGraphClick(TObject *Sender) {
 if (flIgnoreChangeEvent) return;
 if (TabControllVariants->TabIndex<0) return;       // - ???

 InputValues *pIV = (InputValues *) TabControllVariants->Tabs->Objects[TabControllVariants->TabIndex];
 pIV->iDrawGraph = CheckBoxDrawGraph->Checked;

 if (CheckBoxDrawGraph->Checked) {
        if (!pIV->pResults) {
                InputValuesGetFromForm(pIV);
                Recalc(pIV);
```

```cpp
            SetCaptions(pIV->pResults);

            if (!pIV->pResults) return;

    }

    if (!pFormGraph) pFormGraph = NewFormGraph(); // new TFormGraph(this);

}

if (pFormGraph) {

    pFormGraph->UpdateGraphData();

    //pFormGraph->BringToFront();

}

}

//---------------------------------------------------------------------------

void __fastcall TFormMain::FormCloseQuery(TObject *Sender, bool &CanClose) {

 if (!isModifyVariants) return;

 // Confirm:

 AnsiString msg = "Save changes to file\n\"" + sFileVariants + "\"" ;

 int ret = Application->MessageBoxA(msg.c_str(), NULL, MB_YESNOCANCEL|MB_ICONQUESTION); // MB_ICONWARNING

 if (ret==IDNO) return;

 if (ret==IDYES) {

    VariantSaveToFile(sFileVariants.c_str());

 }

 if (ret==IDCANCEL || isModifyVariants) {
```

```
        CanClose = false;

        return;

    }
}

//---------------------------------------------------------------------------

const Results *GetResults(int &indResults, bool * pflActive, AnsiString *psVariantIdent,
AnsiString *psVariantComment) {

 int nV = FormMain->TabControllVariants->Tabs->Count;

 for(int i=indResults; i<nV; i++) {

        const InputValues *pIV = (const InputValues *) FormMain->TabControllVariants->Tabs->Objects[i];

        if (!pIV->iDrawGraph || !pIV->pResults) continue;

        indResults = i;

        if (pflActive)

                *pflActive = i == FormMain->TabControllVariants->TabIndex;

        if (psVariantIdent)

                *psVariantIdent = pIV->sVariantIdent;

        if (psVariantComment)

                *psVariantComment = pIV->sVariantComment;

        return pIV->pResults;

 }
 return NULL;

}

//---------------------------------------------------------------------------
```

```cpp
void __fastcall TFormMain::BtnViewLogFileClick(TObject *Sender) {
 if (TabControllVariants->TabIndex<0) return;        // - ???
 const InputValues *pIV = (const InputValues *) TabControllVariants->Tabs->Objects[TabControllVariants->TabIndex];
 if (!pIV->pResults) {
        BtnViewLogFile->Enabled = false;
        return;
 }

 if (pIV->pResults->pFormViewLog) {
        if (pIV->pResults->pFormViewLog->WindowState==wsMinimized)
                pIV->pResults->pFormViewLog->WindowState = wsNormal;
        else    pIV->pResults->pFormViewLog->Show();
 } else {
        pIV->pResults->pFormViewLog = new TFormViewLog(this);

        AnsiString sVariant = pIV->sVariantIdent + " ("+pIV->sVariantComment+")";
        pIV->pResults->pFormViewLog->Caption = "Log: " + sVariant;
        pIV->pResults->pFormViewLog->Memo1->Text =
                "// File:\t\"" + ExpandFileName(sFileVariants) +
                "\"\r\n// Variant:\t" + sVariant + "\r\n\r\n" + pIV->pResults->sLogInfo;
        pIV->pResults->sLogInfo = "";

        pIV->pResults->pFormViewLog->RecommendedFileName =
ChangeFileExt(ExpandFileName(sFileVariants), "."+pIV->sVariantIdent+".txt");
        pIV->pResults->pFormViewLog->Show();
```

}

}

//-------------------------------------------------------------------------

#ifdef _DEBUG

#endif

//-------------------------------------------------------------------------